\documentclass[12pt,preprint]{aastex}
\usepackage{graphicx}

\shorttitle{Blue-shifted loop}
\shortauthors{Durgesh Tripathi et al.}

\begin{document}

\title{Observations of Plasma Upflow in a Warm Loop with Hinode/EIS}
\author{Durgesh Tripathi}
\affil{Inter-University Centre for Astronomy and Astrophysics, Post Bag 4, Ganeshkhind, Pune 411 007, India}

\author{Helen E. Mason and Giulio Del Zanna}
\affil{Department of Applied Mathematics and Theoretical Physics, University of
Cambridge, Wilberforce Road, Cambridge CB3 0WA, UK}

\author{Steven Bradshaw}
\affil{Department of Physics and Astronomy, Rice University, Houston, TX 77005, USA}

\begin{abstract}A complete understanding of Doppler shift in active region loops can help probe the basic physical mechanism 
involved into the heating of those loops. Here we present observations of upflows in coronal loops detected in a range of 
temperature temperatures ($\log\, T=5.8 - 6.2$). The loop was not discernible above these temperatures. The speed of upflow 
was strongest at the footpoint and decreased with height. The upflow speed at the footpoint was about  20~km~s$^{-1}$ in \ion{Fe}{8} 
which decreased with temperature being about 13~km~s$^{-1}$ in \ion{Fe}{10},  about 8~km~s$^{-1}$ in \ion{Fe}{12} and 
about 4~km~s$^{-1}$ in \ion{Fe}{13}. To the best of our knowledge this is the first observation providing evidence of upflow of plasma in 
coronal loop structures at these temperatures. We interpret these observations as evidence of chromospheric evaporation in quasi-static 
coronal loops.
\end{abstract}


\section{Introduction}

The problem of heating and maintaining the structures in the upper solar atmosphere, such as transition region and corona, 
has been considered to be one of the most challenging issues in astrophysics. There have been tremendous developments in 
observations and theory in the past few decades. However, a definitive solution remains elusive. See \cite{klimchuk_2006} for 
a review.

The discovery that loop structures are one of the basic building blocks of the solar corona has simplified the problem a great 
deal to that of understanding the heating and maintenance of individual loops. This is basically because of the inherent 
characteristics of the corona are such that there is no cross field conduction due to very high electric conductivity.

Recent observations have revealed that active regions comprise different types of loop structures: warm loops (which seem to 
have a temperature around 1~MK), fan loops (seen emanating from the umbral and penumbral regions at temperatures slightly 
lower than 1~MK), and hot core loops (seen in the core of active regions at temperatures around 2$-$3~MK). The heating and 
maintenance of fan loops and hot core loops are currently a matter of hot debate. For hot core loops, there is evidence for both 
steady \citep[see e.g.,][]{warren_2008, brooks_warren, warren_2010, tmdy_2010, wine_2011} as well as impulsive 
heating \citep[][]{tmk_2010, tkm_2011, vk_2012, tmk_2012}. However, the observed characteristics of warm loops such as 
density, filling factors and temperature distribution appear to be consistent with those derived from an impulsive heating model 
\citep[see e.g.,][and references therein] {warren_2003, tripathi_2009, inaki_2009, klimchuk_2009}.

The Doppler shift measurements in previously reported observations of warm loops show predominantly redshifted emission 
along the loop structures, at around 1~MK \citep[see e.g.][]{delzanna_2008, tripathi_2009}. At lower temperatures
however, the red shifts are localised towards the footpoints and stronger in magnitude. These downflows (redshifts) are 
interpreted as plasma radiatively cooling and condensing in the loops \citep[see e.g. ][]{steve_2008, steve_2010}. 
However, the question remains with regard to how this plasma rises up into the loops at first place. 

The impulsive heating scenario predicts that the plasma flows up into these loops via the mechanism of chromospheric 
evaporation which occurs at relatively higher temperatures. \cite{patsourakos_2006} predict that those high velocity upflows will 
be clearly seen in \ion{Fe}{17} lines, or other lines forming at similar temperatures. However, so far there has been no direct 
detection of these upflows. Such upflows could also be attributed to Type-II spicules \citep[see e.g.][]{spicule2}.

Here we present the first observations of a warm loop which shows
clear upflows of the plasma at low temperatures. These observations are taken with 
the Extreme-ultraviolet Imaging Spectrometer \citep[EIS;][]{eis} abroad Hinode. The paper is structured as follows. In 
section~\ref{obs} we describe the observations followed by data analysis and results in section~\ref{res}. We summarise and 
conclude in section~\ref{summ}.

\section{Observations} \label{obs}

The Extreme-ultraviolet Imaging Spectrometer \cite[EIS;][]{eis} onboard Hinode observed an active region, AR 11131, on 
11-Dec-2010 using the study sequence which we designed ('GDZ\_300x384\_S2S3\_35'). This sequence uses the 2~\arcsec slit with an exposure 
time of 35 seconds. This study sequence is designed to be a sparse
raster with a step size of 3~\arcsec. The EIS raster used in 
this analysis started at 01:55:26 UT and was completed at 02:56 UT. We have applied standard processing software provided 
in \textsl{solarsoft} to the EIS observations.

\begin{figure}
\centering
\includegraphics[width=1.0\textwidth]{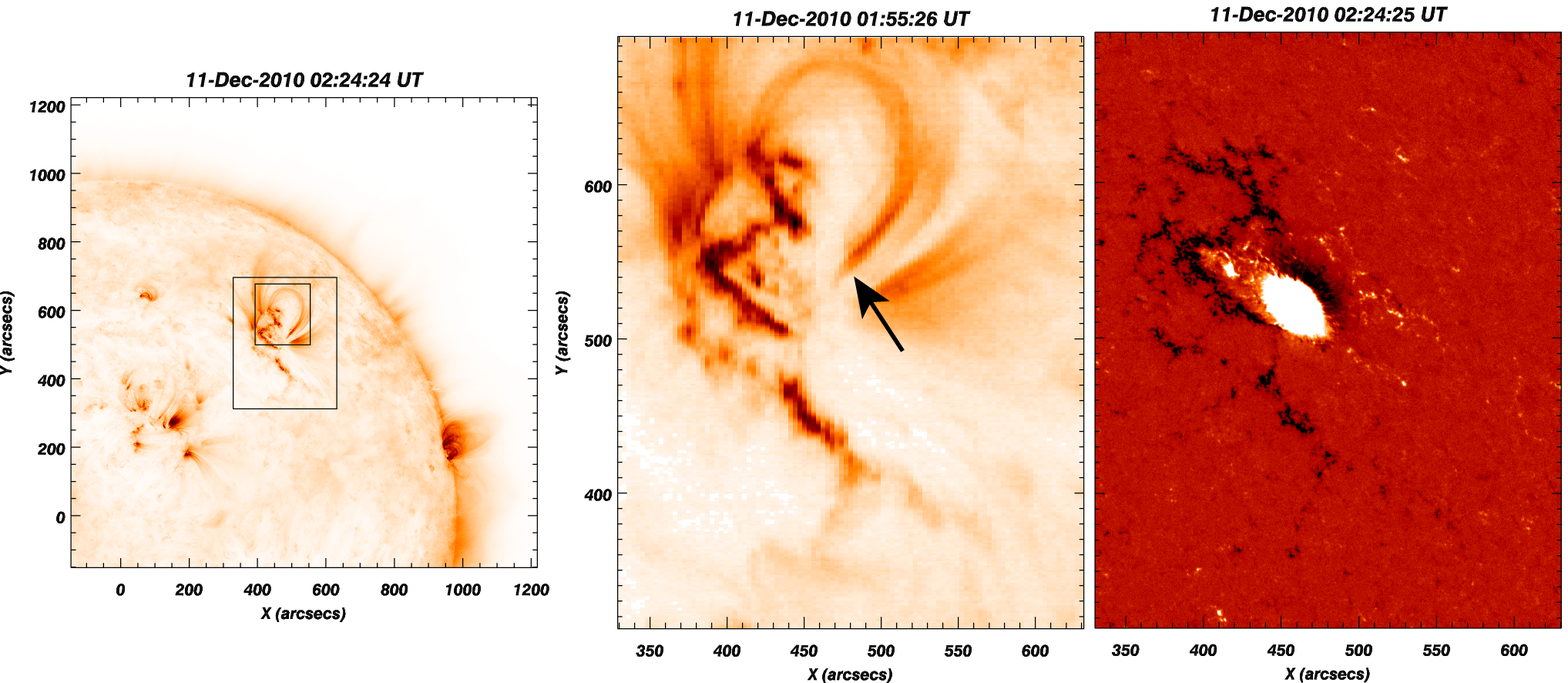}
\caption{Left panel: AIA 171~{\AA} image. The bigger box shows the EIS field of view and smaller box show the region which is 
studied in detail in Fig.~\ref{aia_zoom}. Middle Panel: EIS \ion{Fe}{10} image. The arrow locates the foot point of the loop structure 
which is subject of the study. Right Panel: HMI line of sight magnetic field corresponding to EIS FOV. The magnetic field is scaled 
between -300 and 300 Gauss. \label{aia_eis}}
\end{figure}

The left image in Fig.~\ref{aia_eis} displays a part of the Sun's disk recorded by the 171~{\AA} passband of Atmospheric 
Imaging Assembly (AIA) onboard the Solar Dynamics Observatory (SDO). The emission in 171~{\AA} passband 
peaks at a temperature of log~T=6.0, similar to the \ion{Fe}{10}~184~{\AA} line observed by the EIS. However, 
depending on the individual structures being observed on the Sun, different channels on AIA may respond differently 
\citep[see e.g.,][]{brendan_aia, delzanna_2011}. The over-plotted bigger box shows the area which was rastered by EIS, for which a 
spectral image built in the \ion{Fe}{10}~184~{\AA} is shown in the middle panel. The smaller box in the left  panel represent the region which was 
further studied in detail by AIA. The right panel shows the line-of-sight magnetogram obtained by the Helioseismic and 
Magnetic Imager (HMI) aboard SDO for the region corresponding to the
EIS raster. In the magnetogram, white regions represent the positive magnetic field and dark represent the negative magnetic field regions. 
AIA and HMI images are taken when the EIS slit would have been approximately 
at the middle of the raster. It is worthwhile to note that we have coaligned EIS and AIA images just by eye. We believe that 
the co-alignment is accurate to a level of about 7-8 arc sec. The loop structure, on which this paper is focussed is located by an arrow. 
These images clearly show that the western footpoint of the loop is rooted in the sunspot seen in the right panel.
The eastern moss region corresponding to negative polarity regions is more dispersed. It is worthwhile to note that the moss regions 
are seen only towards the negative polarity region as noted earlier by \cite{tripathi_2008}.

\begin{figure}
\centering
\includegraphics[width=0.8\textwidth]{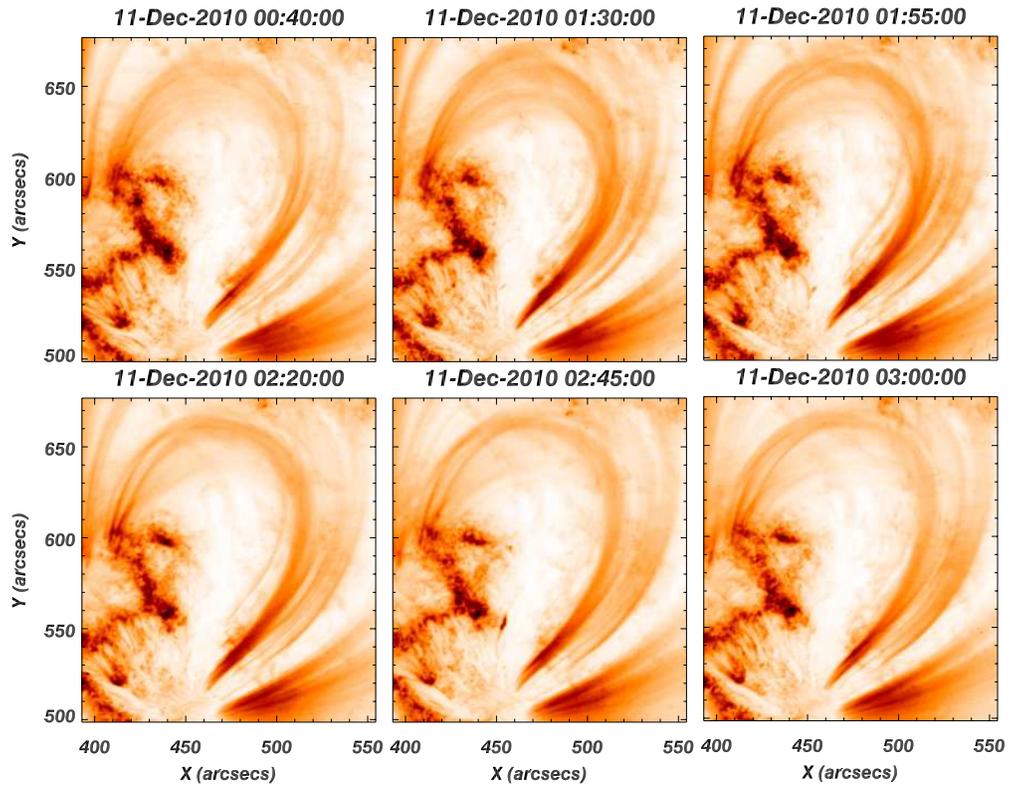}
\caption{AIA 171~{\AA} images corresponding to the small box shown in Figure~\ref{aia_eis}. The images are shown before, during and after the EIS raster.
\label{aia_zoom}}
\end{figure}

Since we are interested in an understanding of quiescent coronal loops, we analysed the AIA data for about 5 hours 
(a couple of hours before and after the observations) to make sure that the loop being analysed here was quasi-static 
and that the active region did not show any flaring and/or micro-flaring activity. Figure~\ref{aia_zoom} displays AIA images 
recorded in the 171~{\AA} band corresponding to the smaller box shown in the left panel of Fig.~\ref{aia_eis}. It is clear from 
these images that there are no major changes in the overall structures of the loops. However, some small variations in 
the intensity towards the footpoint region can be seen. The results presented here relate to quiescent (1MK)
coronal loops seen with TRACE or SDO/AIA at 171~\AA.

\section{Data Analysis and Results} \label{res}

Figure~\ref{int} shows intensity maps derived in  \ion{Mg}{5}~276~{\AA}, \ion{Mg}{7}~278~{\AA}, \ion{Mg}{7}~280~{\AA}, 
\ion{Fe}{8}~186~{\AA}, \ion{Fe}{10}~184~{\AA}, \ion{Fe}{11}~188~{\AA}, \ion{Fe}{12}~195~{\AA}, 
\ion{Fe}{13}~202~{\AA} and \ion{Fe}{15}~284~{\AA}. As can be seen from these 
images, the loop structures (shown with arrows in middle left panel) are clearly discernible in \ion{Mg}{7}~278~{\AA}, 
\ion{Si}{7}~275~{\AA}, \ion{Fe}{9}~188~{\AA}. Only a tiny portion of the loop at footpoint, labelled with an arrow in top 
row middle panel, is seen in \ion{Mg}{5} line. The footpoint of the loop seen in \ion{Mg}{5} image is thinner than the same part of 
the loop seen in other lines such as \ion{Si}{7} and \ion{Fe}{9}. For the higher temperature lines such as \ion{Fe}{10} the contrast 
between loop and the background/foreground fuzzy emission has decreased and it becomes almost impossible to discern the 
loop in \ion{Fe}{13}.  In other words the corona becomes fuzzier with increasing temperature as was shown with 
Hinode/EIS data by \cite{tripathi_2009}. It is probably worth emphasising that the intensity in
the loop structures are just about 10-20\% higher than the background/foreground coronal intensities 
\citep[see e.g.][]{giulio_helen, vk_2012}. It is important to note that the intensities in the background/foreground regions are higher in the images 
obtained in lines like \ion{Fe}{12} and \ion{Fe}{13}, but decreases in \ion{Fe}{15}. These images also suggest that there are loops at different 
temperatures intermingled together which can only be distinguished uniquely by spectroscopic observations.

Comparing the structures in AIA images (Fig.~\ref{aia_zoom}, spatial resolution $\approx$1~arcsec) and with that in EIS images 
(Fig.~\ref{int}, spatial resolution $\approx$3$-$4~arcsec) it is clear that the loop structure which appear to be an individual structure in 
EIS images, comprise several different flux tubes as seen with AIA. A comparison between AIA, EIS and HMI observations reveals that 
the western footpoints of the loop structures under consideration are rooted in the sunspot (leading polarity) and the eastern footpoints are rooted in more fragmented following polarity magnetic field regions.

The footpoint of the loop, shown by an arrow labelled as 'F' in the left panel in the middle row of the Fig.~\ref{int}, is brightest. 
With increasing height, the emission from the loop gets fainter. A closer look at the footpoint region suggests that there are at 
least three different loops, marked with three arrows higher up, emanating from either the same location or close by. The 
footpoint regions of these three loops seems to coincide along the line of sight which could 
possibly lead to brighter emission near the footpoints.

The two right panels in the top row in Fig.~\ref{int} show intensity maps obtained using two lines of \ion{Mg}{7} namely 
278~{\AA} and 280~{\AA}. The loop is clearly visible in \ion{Mg}{7}~278~{\AA}, but only a part of the loop is visible in 
\ion{Mg}{7}~280~{\AA}. We note that the ratio of
\ion{Mg}{7}~280~{\AA} to \ion{Mg}{7}~278~{\AA} is sensitive to electron number density. This indicates
that the electron density is very low towards the upper part of the loop. 

Deriving Doppler shifts from EIS observations is a non-trivial as well as a non-unique process \cite[see e.g.][]{kamio_2010, peter_flows, 
tmk_2012}. Most important of all is to remove the orbital variation of the spacecraft and to determine the Ôreference wavelengthÕ. Doppler 
shifts can then be derived. There are a few different ways to derive Doppler shifts from EIS data \citep[see e.g.,][]{kamio_2010, dadashi, 
peter_flows}. In this paper we have used the method proposed by \cite{peter_flows}. This is based on deriving velocities (using \ion{Fe}{8} as a 
reference) in the quiet Sun region of each slit position in the raster. Following this method, we have used the \ion{Fe}{8}~186.6~{\AA}
line in the Quiet Sun to derive Doppler shifts for other lines. We have chosen the bottom 30 rows in the raster be the quiet Sun. The only 
difference between our method and that of \cite{peter_flows} is that they calibrated the quiet Sun using the SUMER measurements from Peter 
and Judge (1999), whereas we choose the \ion{Fe}{8} line to have zero Doppler shift, consistent with the new values of \cite{dadashi}. For more 
details on this method the reader is referred to \cite{peter_flows}.

Figure~\ref{velo} shows the velocity maps derived for \ion{Fe}{8}, \ion{Fe}{10}, \ion{Fe}{12} and \ion{Fe}{13}. The overall velocity structure in 
active region appear to be the similar to those published earlier \citep[see e.g.][]{doschek, delzanna_2008, tripathi_2009}, with weakly redshifted 
core of the active region and strongly blue-shifted low intensity regions. The core of the active region show
strong redshifts for the \ion{Fe}{8} line. This is because the 
\ion{Fe}{8}~186~{\AA} line is blended in its redwing with a \ion{Ca}{14} line which forms at log~T = 6.4. Since the core of the active 
region is characterised by high temperature emission, the Doppler shift for \ion{Fe}{8} is overestimated.
The loop structures marked by arrows in Fig.~\ref{int} are blue shifted, unlike the earlier reported observations of strongly redshifted warm loops. 
The blue shifted loops are marked by arrows in Fig.~\ref{velo}. The
loop which is seen along its entire length in blue shifted all along 
the loop in \ion{Fe}{8} line and both the foot points are visible. For other spectral lines shown in Fig.~\ref{velo}, only one foot 
point of the loop is visible and is blue shifted.

\begin{figure}
\centering
\includegraphics[width=1.0\textwidth]{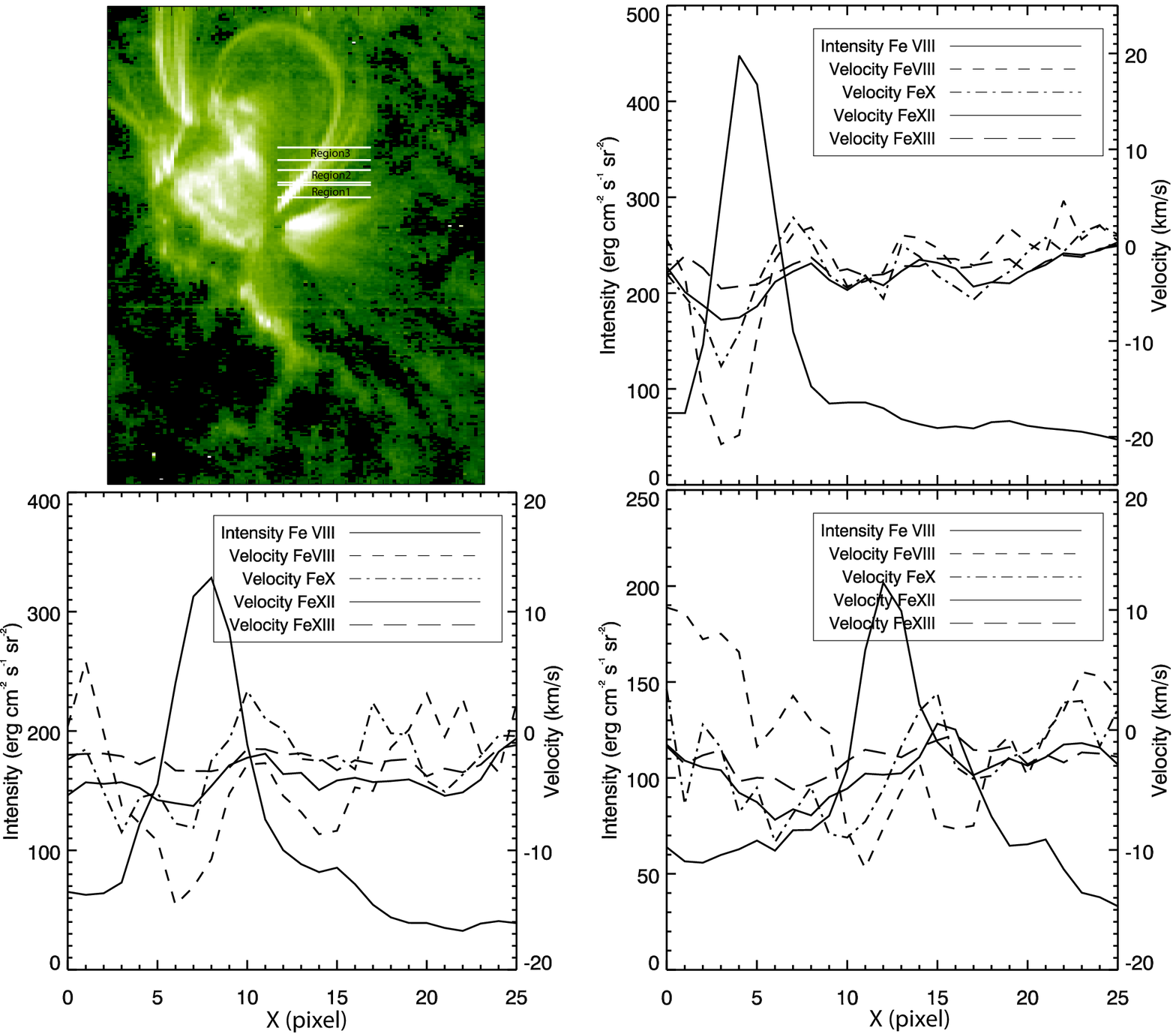}
\caption{Correspondence between \ion{Fe}{8} intensity profile and velocity profiles for \ion{Fe}{8}, \ion{Fe}{10}, \ion{Fe}{12} and \ion{Fe}{13}. \label{int_vel_profile}}
\end{figure}

To understand the correlation between the loop intensity and its velocity structure better, we plot, intensity and velocity profiles along 
three different horizontal cuts shown in top left panel of Fig.~\ref{int_vel_profile} as 'Region~1', 'Region~2' and 'Region~3'. The rest three 
panels show the intensity and velocity profiles for three regions. These intensity is plotted for \ion{Fe}{8} line and the corresponding 
velocity profiles are plotted for \ion{Fe}{8}, \ion{Fe}{10}, \ion{Fe}{12} and \ion{Fe}{13}. In the plots, the strong spikes (intensity enhancement) in 
the intensity profile correspond to the loops marked in Fig.~\ref{int}. As anticipated the loop intensity decreases with height. Similar 
characteristics is seen in the velocity profile. The loop is strongly blue shifted at the foot point with a speed of about 
20~km~s$^{-1}$ in \ion{Fe}{8}. The velocity of the upflow at the foot point decreases with temperature being about 13~km~s$^{-1}$ in 
\ion{Fe}{10},  about 8~km~s$^{-1}$ in \ion{Fe}{12} and about 4~km~s$^{-1}$ in \ion{Fe}{13}. In addition, the upflow velocity also decreases with the 
height of the loop. We note that these speeds are lower limits of the actual speed of the plasma, since we are measuring line-of-site 
component of speeds. Similar, though less pronounced, characteristics can be seen for other loops adjacent to the strongly emitting loop 
structure.

The peak of the intensity and velocity profiles for the loops do not coincide with each other. The peak of the velocity 
seem to be shifted 3{\arcsec} eastward with respect to the peak in intensity. This is most likely due to the tilted point spread 
function of EIS as was pointed out  by \cite{peter_flows}. \cite{peter_flows} found that strongly redshifted or blue shifted regions 
are often seen with some offset from the regions with highest intensity point. Investigating very many different datasets, they 
concluded that this could be attributed to a tilted point spread function of the instrument. In one observation presented in 
\cite{peter_flows} it was found that there was a difference between intensity peak and velocity peak of 3{\arcsec} in Y-direction. In these 
circumstances, \cite{peter_flows} state, "the simplest way of deriving velocity shifts  for features with steep intensity gradients 
is either to study only those pixels at the location of the intensity peaks in the solar-Y direction, or to perform averaging in the 
Y- direction over a region symmetrically distributed around the intensity peak." In the present observations, even if we average 
the velocity over the width of the loop in X-direction, the blue shifted pattern remains. Therefore, we conclude that this 
blue shift seen along the loop is real and not due to some instrumental artifact.

\section{Summary and Discussion} \label{summ}

\begin{figure}
\centering
\includegraphics[width=0.8\textwidth]{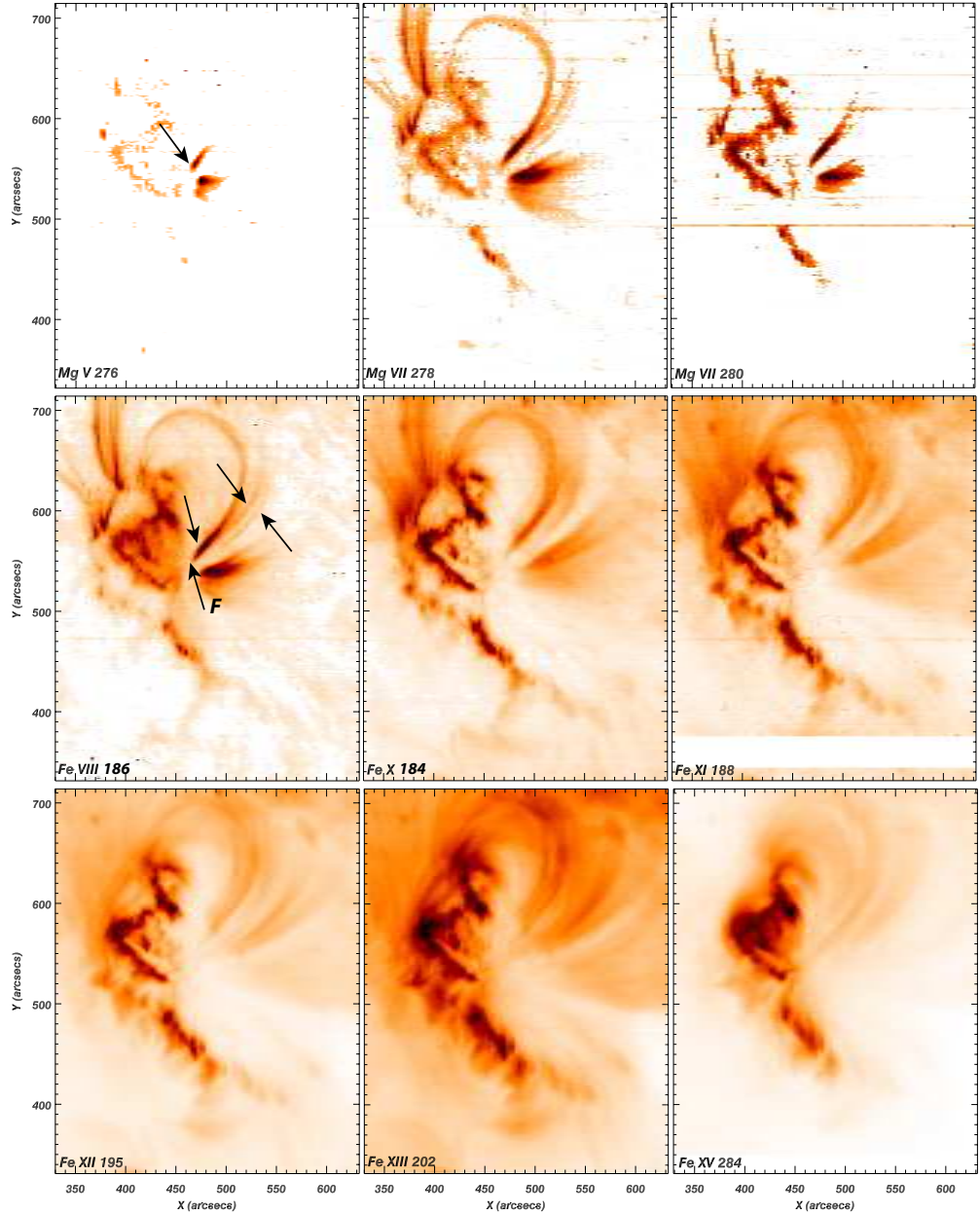}
\caption{Monochromatic intensity maps obtained using EIS spectra. The arrow shown in the top left panel shows the footpoint of the loop 
seen in \ion{Mg}{5} which is also shown by an arrow in top right panel labelled as 'F'. The three arrows shown in the top right panel locate 
three distinguished loops.\label{int}}
\end{figure}

\begin{figure}
\centering
\includegraphics[width=1.0\textwidth]{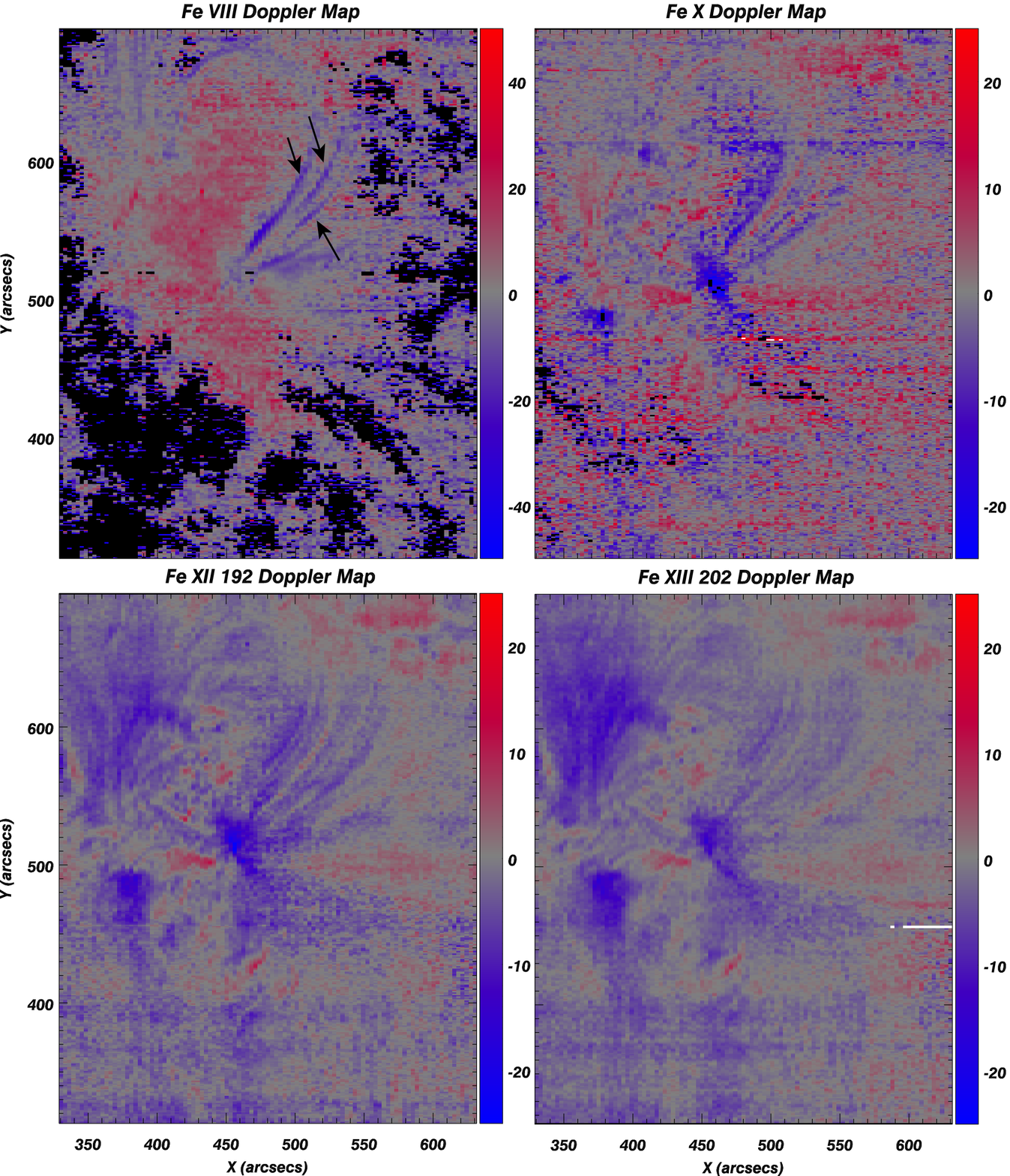}
\caption{Doppler maps for \ion{Fe}{8}, \ion{Fe}{10}, \ion{Fe}{12} and \ion{Fe}{13}. \label{velo}}
\end{figure}

We have presented an observation of the active region AR 11131 recorded on 11-Dec-2010 by Hinode EIS. The active region comprises of 
warm loops, fan loops and hot core loops among other features such as moss etc. The intensity and velocity structure of the active region shows 
similar characteristics as published earlier by e.g. \cite{doschek, delzanna_2008, tripathi_2009}. However, in contrast to the observations 
reported earlier, where the warm loops at around 1~MK have been shown to be redshifted, the warm loops in this active region are blue shifted 
(see Figs~\ref{int} and \ref{velo}). The footpoint of the loop is strongly blue shifted with a speed of about 20~km~s$^{-1}$ in \ion{Fe}{8}. The 
upflow velocity at the foot point decreases with increasing temperature as well as the height of the loop.To the best of our knowledge this is the 
first observational report of a blue shifted warm loop.

Based on the multi-stranded impulsive heating scenario, chromospheric evaporation of plasma into coronal loop structures takes place at rather 
very high temperatures:  \cite{patsourakos_2006} predict high velocity upflows in spectral lines like \ion{Fe}{17}. In this scenario, the heating 
takes place in the corona. Energy is conducted downwards leading to chromospheric evaporation into the corona. At the temperature around 
which lines like \ion{Fe}{8} would form, most of the strands would be cooling down and therefore the loop emission is expected to be dominated 
by redshifted emission. So the observations presented here do not seem to show the flow structure expected from this multi-stranded loop 
models. However, it is important to emphasise here that in these models plasma flow characteristics changes quite significantly when the 
characteristics of the heating events are changed. The characteristics of the heating events are ad hoc. If the energy of the
heating events is significantly higher, then it may well be that high velocity upflows will be seen at higher temperatures and later on the total 
emission will be dominated by plasma condensation. However, if the energy and duration of heating event is short, the plasma may behave 
differently and the model may predict upflows at significantly lower temperature, making our observations consistent with an impulsive heating 
scenario. However, this scenario predicts increasing upflow velocity with temperature, which is opposite to what we observe here.

Our observation is also consistent with the picture that the heating takes place at the foot points of the loops i.e., lower down in the atmosphere 
rather than higher up in the corona \citep[see e.g.][]{foot_heating}. In this scenario too, because of the heat deposition lower down will 
increase the pressure leading to the plasma evaporation into the loops. If the heat deposition is lower in the atmosphere then evaporation at 
temperatures of about 1~MK would result in blue shifted emission in the loops, similar to our observations. In fact this is argued to be one 
of the best evidences for footpoint heating \citep[see e.g.][]{foot_heating}.

Lastly, the observations presented here could possibly be consistent with newly proposed mechanism of heating of the solar corona by spicules 
type II \citep[][]{spicule2}. In this scenario, the chromospheric material is being pumped into the corona by these spicules and heated. 

To the best of our knowledge, these observations present the first clear case of the upflow of plasma in coronal loops and provide an important 
constraint on the theories of coronal heating. Clearly more observations and theoretical work are needed before anything conclusive can be 
said about the heating of warm loops. The future missions such as the Interface Region Imaging Spectrometer (IRIS), the Spectral Imaging of 
Coronal Environment on board Solar Orbiter and the proposed Large European module for solar Ultraviolet Research (LEMUR) onboard 
Solar-C will provide with unique and definitive observations to study such problems.

\acknowledgments{We thank the referee for providing useful inputs which has improved the paper. HEM and GDZ 
acknowledge STFC. We acknowledge useful discussions at the ISSI on Active Region Heating. Hinode is a Japanese mission developed 
and launched by ISAS/JAXA, collaborating with NAOJ as a domestic partner, NASA and STFC (UK) as international partners. Scientific 
operation of the Hinode mission is conducted by the Hinode science team organized at ISAS/JAXA. This team mainly consists of scientists 
from institutes in the partner countries. Support for the post-launch operation is provided by JAXA and NAOJ (Japan), STFC (U.K.), NASA, 
ESA, and NSC (Norway). }

\end{document}